\documentclass[reprint,aps,amsfonts, amssymb, amsmath, prb, showkeys]{revtex4-2}

\usepackage{dcolumn}
\usepackage{bm}
\usepackage{array,multirow}
\usepackage{adjustbox}
\usepackage{graphicx}

\begin{document}

\title{Non-local origin and correlations in the Johnson noise at nonuniform temperatures}

\author{Jorge Berger}
\address{Department of Physics, Braude College, Karmiel, Israel\footnote{jorge.berger@braude.ac.il\\
\url{https://orcid.org/0000-0002-1900-6707}}}

\author{Guy Katriel}
\address{Department of Mathematics, Braude College, Karmiel, Israel}

\begin{abstract}
We propose an alternative scenario for the propagation of thermal noise in a conductor. In this scenario, the noise in the emf (electromotive force) between two terminals cannot be described as a sum of contributions from uncorrelated regions, each in local thermal equilibrium. We review previous studies of thermal noise in circuits with nonuniform temperature. We suggest experiments that could distinguish between different scenarios. 
We build a workable 1D model for a gas of particles that undergo stochastic collisions with the lattice and exert distance-dependent forces on each other.
We enunciate definitions of current, voltage, and emf, appropriate to a wire with limited number of particles. For uniform temperature, within appropriate length and temperature ranges, our simulations comply with Nyquist's result.
Analytic results can be obtained in the limit of strong interparticle interaction. The simulations indicate that (1) thermal noise in a resistor at uniform temperature within an electric circuit can be larger (smaller) than predicted by Nyquist due to the presence of a resistor with higher (lower) temperature in the circuit; (2) for sufficiently long circuits, the deviation from the Nyquist prediction is inversely proportional to the distance between the centers of the resistors; (3) if the resistors differ in temperature, their emf can be correlated, even if they are detached. 
The long-range repulsion between charges in electrically connected resistors may have conceptual and technological impact in nanodevices.

\end{abstract}

\maketitle

\section{Introduction}
The electromotive effect of the thermal agitation of the carriers of electricity in conductors was predicted by Einstein,\cite{ein} measured by Johnson,\cite{john} and explained by Nyqist \cite{ny} by invoking the second law of thermodynamics and detailed balance. This effect is called ``Johnson noise'' and will be spelled out below. 

Noise sets a limitation to every measurement, unless it can be performed at sufficiently low temperatures. On the other hand, in recent decades, noise has become a measuring tool,\cite{Qu} and the assumption that thermal noise depends only on the resistance and the temperature allows its use as an ideal thermometer. Due to its wide range of applicability, Johnson noise thermometry is the tool of choice for many applications and characterization of materials, but also for fundamental questions, such as redefinition of the Kelvin \cite{KB1,KB3} or observation of fractional charges in the quantum Hall effect.\cite{frac,shot}

It is customary to describe Johnson noise in the frequency domain, but for our purposes, it will be clearer to describe it in the time domain: when a current $I$ is driven  along a resistor with resistance $R$ at temperature $T$ and the voltage is measured during a lapse of time $\Delta t$, the time-average voltage drop across the resistor is not just $IR$, but, due to the Johnson noise, it is $IR+\Delta V$, where the ensemble average of $\Delta V$ is 0 and its variance is \cite{Bakker}
\begin{equation}
    \langle (\Delta V)^2\rangle =2k_BTR/\Delta t
 \label{varV}\,,
\end{equation}
where $k_B$ is the Boltzmann constant. For very small values of $T$ or $\Delta t$ there are deviations from (\ref{varV}), that are beyond the scope of this study. Likewise, if a voltage $V$ is applied to the resistor, the electric current will have a time-average $V/R+\Delta I$, where the ensemble average of $\Delta I$ is zero and its variance is
\begin{equation}
    \langle (\Delta I)^2\rangle =2k_BT/R\Delta t
 \label{varI}\,.
\end{equation}
The derivation of these results is discussed in Appendix \ref{Brown}.

In this study we will consider circuits without power sources, and in the following we will drop the $\Delta$ from $\Delta V$ and $\Delta I$. Equations (\ref{varV})-(\ref{varI}) are applicable to particular situations in which either the current or the voltage is fixed and does not fluctuate; in the general case (e.g. in a segment of a wire) the Johnson noise is described by means of the emf (electromotive force) $\varepsilon =V+IR$. Its variance is
\begin{equation}
    S(\varepsilon ):=\langle \varepsilon^2\rangle - \langle \varepsilon\rangle^2=2k_BTR/\Delta t
 \label{varep}\,.
\end{equation}

The physical agent responsible for this noise is a fluctuating electric field in addition to that applied by power sources. 
The question we intend to address is: Where does this field originate? The naive answer is that this field is a local effect, induced by thermal motion of the charged material that makes up the resistor.

This local-origin view was present in Nyquist's description. Reference \cite{ny} regards a resistor as a power source, with the power originating inside it, as expressed in passages such as ``The electromotive force due to thermal agitation \textit{in} conductor I...'' or ``...the member in which the electromotive force is generated.'' We will call this view the ``Nyquist scenario'' (NS).

\section{Motivation for an alternative scenario}
\subsection{Heuristic arguments}
We know that an electric signal propagates along a wire or, in general, through a conductor. From a detailed perspective, this propagation could be described as a plasma-electromagnetic wave, subject to the condition that no current crosses the boundary of the conductor. We conjecture that the electric field generated by thermal agitation is not an exception, and can therefore propagate along an electric circuit. Accordingly, we envisage the following scenario: thermal motions \textit{inside all the conducting material that makes up the circuit} create or modulate plasma-electromagnetic waves that pervade the entire circuit and give rise to the voltages or currents in the Johnson noise.

\begin{figure}
\scalebox{0.5}{\includegraphics{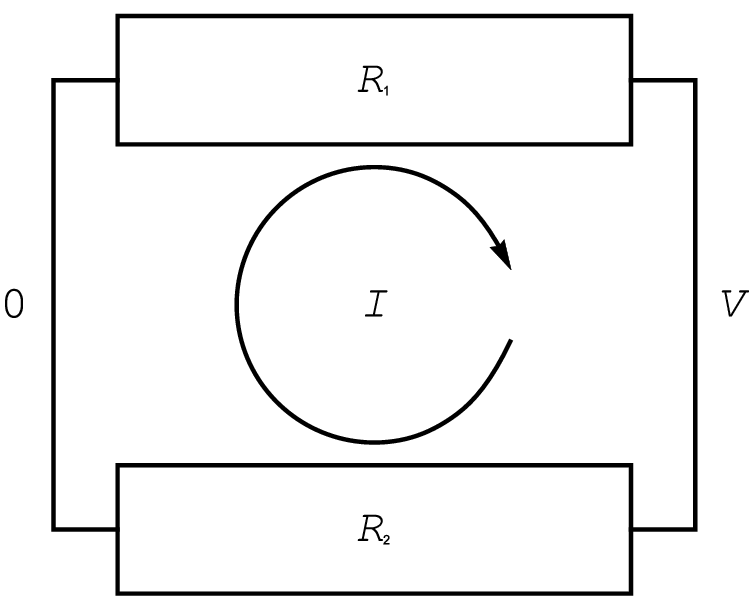}}
\caption{\label{2res}A circuit with two resistors. There are no power supplies in the circuit, but, due to thermal noise, there is a voltage drop (defined as positive at the right end) and a current (positive if clockwise).}
\end{figure}

As a simple example, we may consider a closed circuit with two resistors, $R_1$ and $R_2$, as sketched in Fig.\ \ref{2res}. We can regard the circuit as a pair of resistors in parallel, with equivalent resistance $R_1R_2/(R_1+R_2)$, with no net current flowing from one end of the equivalent resistor to the other. Elaborating in (\ref{varV}), it follows that the time-average voltage across the ends of the equivalent resistor during a lapse of time $\Delta t$ is
\begin{equation} V=\eta_V\sqrt{2k_BTR_1R_2/(R_1+R_2)\Delta t} \,,
\label{etaij}
\end{equation}
where $\eta_V$ is a random variable with average 0, variance 1, and normal distribution. 

We can also regard the circuit as a pair of resistors in series, with an equivalent resistance $R_1+R_2$ and no voltage drop across the equivalent resistor. Therefore, it follows from (\ref{varI}) that the time-average current around the pair of resistors during the lapse of time $\Delta t$ is
\begin{equation} 
I=\eta_I\sqrt{2k_BT/(R_1+R_2)\Delta t} \,,
\label{etaI}
\end{equation}
where $\eta_I$ is a random variable with average 0, variance 1, and normal distribution.
During $\Delta t$, the electric field performs work $VI\Delta t$ on the charges in $R_1$ and $-VI\Delta t$ on the charges in $R_2$, with
$VI\Delta t=2\eta_V\eta_Ik_BT\sqrt{R_1R_2}/(R_1+R_2)$.

We note that in this example each resistor has to ``know'' what is the resistance of the other resistor. The reason it ``knows'' is that signal propagation is built-in into the NS through the assumption that the current is uniform along the circuit. In this example our conjecture is that if a resistor is able to ``know'' the resistance of its companion, it should also be able to ``know'' its temperature. 

\begin{figure}
\scalebox{0.5}{\includegraphics{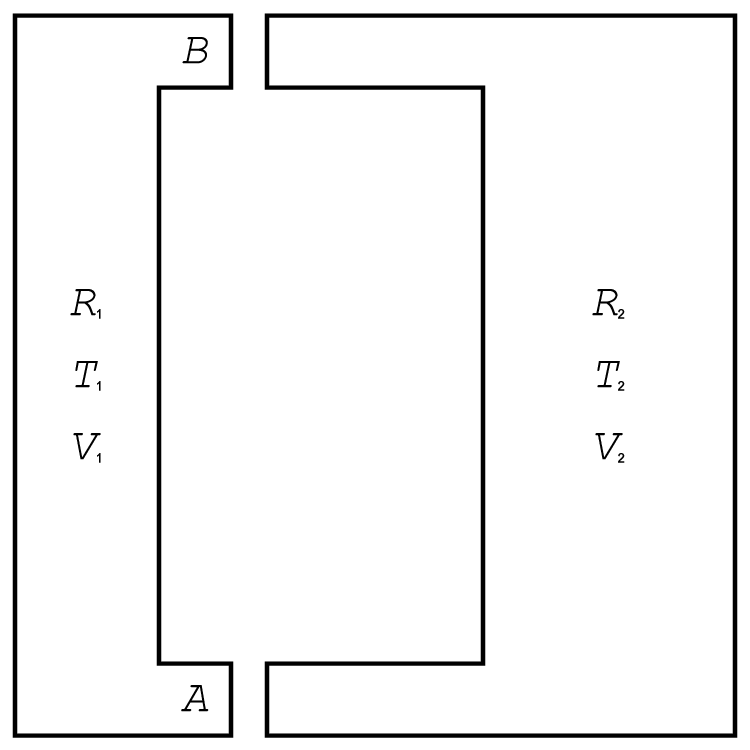}}
\caption{\label{verify} Experiment intended to determine which is the pertinent scenario. Two resistors are initially separated and their temperatures are different. The variance of the voltage between A and B is measured before and after the resistors touch each other and form a closed circuit.}
\end{figure}

The difference between two scenarios is physically irrelevant unless they lead to different falsifiable predictions. We therefore require experiments that can confirm or reject the local or the non-local scenario. 
Since Nyquist's result is based on thermodynamics and detailed balance, any admissible scenario is expected to lead to the same result in situations with uniform temperature; only when more than one temperature is involved, their predictions can differ.

For example, if two resistors are connected and the variance of the emf due to Johnson noise in one of the resistors depends on the temperature of its neighbor, then the NS should be rejected.

As another example, let us consider two initially separate resistors, $R_1$ and $R_2$, with volumes $V_1$ and $V_2$ at different temperatures, $T_1$ and $T_2$, as sketched in Fig.\ \ref{verify}. To fix ideas, let us assume that $R_1\ll R_2$, $V_1\ll V_2$, $T_1\gg T_2$ and $R_1T_1\gg R_2T_2$. Let the variance of the voltage across $R_1$ be measured as reviewed e.g.\ in \cite{Qu}. At some moment, the two resistors are connected. 

Let $\tau_{\rm th}$ be the time required by the resistors in Fig.\ \ref{verify} to approach thermal equilibrium. Then, the effective temperatures of the resistors after, say, $0.1\tau_{\rm th}$ will still be close to the temperatures they had before making contact and, according to NS, $R_1$ would essentially be an unloaded power source with a voltage variance similar to that before contact was made. On the other hand, according to the scenario suggested here, we may expect that the electromagnetic waves have already propagated to the entire circuit and, since $V_1\ll V_2$, they would be close to the case of temperature $T_2$ and the Johnson noise would decrease significantly.

Many variations of this experiment can be conceived. Instead of separate or joined resistors, electrical contact could be controlled by a MOSFET or HEMT transistor. A steady-state version can be obtained if the resistors are permanently connected and a part of each of them is in thermal contact with a heat bath. By periodically connecting and disconnecting the resistors, a lock-in analysis could measure the time lag between the electrical connection and the change in the voltage variance. 

Nonuniform temperature can not only take the form of a function of position, but may also arise as a difference between the temperature $T_e$ of the electrons and the temperature $T_{ph}$ of the lattice. It is well established \cite{Te1,Te2} that, when current is driven along a resistor and the electron-phonon coupling is weak, then $T_e\gg T_{ph}$, and the temperature in Eqs.\ (\ref{varV})-(\ref{varI}) should be replaced with $T_e$. Obviously, in this case the Johnson noise cannot be attributed to ``thermal agitation \textit{in} the conductor'', because there is no such thermal agitation. In this context, our claim can be stated as follows: the strong repulsion between electrons, which nearly imposes electroneutrality, effectively acts as a long-range interaction along the circuit that diverts the electrons from their local equilibrium.

\subsection{Available experimental result}
Strictly speaking, since NS is based on equilibrium thermodynamic arguments, it does not predict anything in the case of nonuniform temperature. However, it can be naturally extended to the case of a wire with steady temperature $T(x)$, $0\le x\le L$. 
From (\ref{varV}), denoting by $\rho$ the resistance per length and assuming that the Johnson noise is a local effect, the variance of the emf (and of the voltage, if there is no current) between $x$ and $x+dx$ would be $2k_BT(x)\rho (x)dx/\Delta t$. Assuming also that the fluctuating voltages in different segments are uncorrelated, we obtain
\begin{equation}
    (\Delta t/2k_B)\langle V^2\rangle =\int_0^L T(x)\rho (x)dx \;.
    \label{steady1}
\end{equation}

Denoting longitudinal average by an overline [e.g. $\overline{T}=L^{-1}\int_0^LT(x)dx$], substituting in (\ref{steady1}) $T(x)=\overline{T}+[T(x)-\overline{T}]$ and $\rho (x)=\overline{\rho}+[\rho (x)-\overline{\rho}]$, and integrating each term separately, (\ref{steady1}) becomes
\begin{equation}
    (\Delta t/2k_B)\langle V^2\rangle =\overline{T}R+L\overline{(T-\overline{T})(\rho -\overline{\rho})} \;,
    \label{steady2}
\end{equation}
where $R$ is the resistance of the wire.

Monnet, Ciliberto and Bellon (MCB) \cite{Bellon} measured the variance of the voltage between the ends of a wire with nonuniform temperature.
In the MCB experiment the wire was made of a metallic alloy, $\rho$ increased linearly with temperature, implying $\overline{(T-\overline{T})(\rho -\overline{\rho})}>0$, and therefore the variance of the voltage was expected to be greater than $2k_BR\overline{T}/\Delta t$ or, translated into MCB notation, the result expected from this extended NS was $T^{\rm fluc}>T^{\rm avg}$. However, as observed for the larger temperature differences in Fig.\ 7 of MCB, they found that $T^{\rm fluc}<T^{\rm avg}$.
$T^{\rm avg}-T^{\rm fluc}$ is just about twice the estimated experimental error, but remains positive throughout the region where the experiment was most sensitive to this difference. A possible explanation could be a negative correlation between different segments, but measurements are available for the entire wire only, and not for segments of it.

Another explanation is that $\overline{(T-\overline{T})(\rho -\overline{\rho})}$ was small (translates into about three times the estimated error in $T^{\rm avg}$ or $T^{\rm fluc}$) and the measurement was unable to detect it. We suggest an alternative explanation: the Johnson noise is nonlocal, and $T(x)$ has to be replaced with a smeared value. Smeared values are closer than the local values to the longitudinal averages, and therefore the effective value of $\overline{(T-\overline{T})(\rho -\overline{\rho})}$ was hindered.
The configuration considered by MCB, in which cold and hot short segments alternated, could have facilitated this smearing effect.

In the MCB experiment, $|T-\overline{T}|\lesssim 0.13\overline{T}$ and $|\rho -\overline{\rho}|\lesssim 0.2\overline{\rho}$, and therefore $L\overline{(T-\overline{T})(\rho -\overline{\rho})}$ is smaller than $\overline{T}R$ by two orders of magnitude. $\overline{(T-\overline{T})(\rho -\overline{\rho})}$ would increase quadratically with the difference in temperature between the hot and the cold reservoir. $|\rho -\overline{\rho}|$ would also be significantly larger for an intrinsic semiconductor with energy gap of the order of $k_B\overline{T}$ or for a conductor with very nonuniform cross section. 

An additional experiment that could have tested the locality of Johnson noise was reported in \cite{prl,Berut}. Two resistors held at different temperatures were connected through a capacitor. Indeed, the variance of the voltage in the hot (cold) resistor was smaller (larger) than predicted by Eq.\ (\ref{varV}) (as follows from Eq.\ (16) in \cite{Berut}). However, this result does not necessarily contradict the NS, because the current along the resistors did not vanish; in this case Eq.\ (\ref{varV}) should not be applied to the voltages, but rather to the electromotive forces (emf), that were not evaluated.

\subsection{Multiterminal conductors}
These are conductors with an arbitrary number of terminals, at which the temperature and either the current or the electric potential are fixed, whereas the rest of their surface is impermeable to charges and heat. Sukhorukov and Loss (SL) \cite{Loss} treated this system following a Boltzmann-Langevin approach. Their starting equation (2.1) accounts for the interaction between electrons as a local scattering agent, whereas the long-range influence of the Coulomb interaction is not considered.

SL predicted that the strength of the noise at a given terminal depends on the \textit{temperatures at the other terminals}, as expressed by their Eqs.\ (3.16) and (6.2) (where $\Pi$ plays the role of a local effective temperature). This prediction is still far from being fully investigated. In the case of junctions (that is, circuits with a place where the temperature jumps and with negligible resistance in the rest of the circuit), it is found \cite{Nature,Ukr} that the variance of the current can be described by a term like (\ref{varI}), with $T$ replaced by the average of the temperatures at both sides of the junction, and an additional term that increases with the temperature jump; other possibilities are analyzed in Ref. \cite{Zhang} and references therein. 

Despite its non-local form, it might be argued that Eq.\ (3.16) in SL provides an extension of Eq.\ (\ref{steady1}) to the case of a three-dimensional conductor in which the temperature is fixed only at the terminals. Rather than addressing this point of view, we will focus on resistive wires with lattice temperature that is defined by tight thermal contact with baths along most of their lengths.

\section{Simulations}
Since the experimental evidence is scarce, we would like to reinforce it by means of model simulations. For our purposes, a ``resistor'' is an object that obeys Ohm's law. If it obeys Eqs.\ (\ref{varV})-(\ref{varI}) for uniform temperature, but does not obey Eq.\ (\ref{steady1}) for nonuniform temperature, we can regard these results as evidence supporting the non-local scenario.

\subsection{Choice of the model\label{sim}}
We would like to mimic the typical features of a conducting wire: charge carriers are free to move but undergo collisions against defects and phonons that lead to their thermalization; these carriers repel each other, thus becoming spread with a nearly uniform density; the repulsion force enables signal propagation at a speed that is faster than the mean carrier speed; the medium in which the carriers move screens long range interactions. In addition to these features, we would like the model to be as tractable as possible.

We model a resistor as a one-dimensional system of $N$ classical particles, each with charge $q$ and mass $m$, in a wire in the region $0\le x\le L$. We define a sort of screening length $\ell$ such that the particles repel each other for distances $r$ smaller than $\ell$ and do not interact for $r>\ell$. For $r<\ell$, the size of the interparticle force is taken as $k_c(\ell-r)$, where $k_c$ is a positive constant. Each particle may also feel a force due to an applied voltage. 

We define a free flight time $\tau$ such that for times that are integer multiples of $\tau$ every particle ``collides'' and the velocity of each particle acquires a random value, with zero average, variance $k_BT/m$, and normal distribution. Between collisions, the motion of the particles is dictated by the forces that act on them. 

 For this modified Drude model \cite{Drude} (the distribution of times between consecutive collisions is sharp), the dc resistance is
 \begin{equation}
 R_D=2mL^2/q^2N\tau\;.
 \label{Drude}
 \end{equation}

 For $k_c\ell^2\ll k_BT$ or $N\ell\ll L$ we expect that the system will behave qualitatively as a gas (collisions act as a heat bath, the interaction between particles has little influence, and particles can pass through each other); in the opposite case, we expect that the system will behave as an incompressible fluid and there will be fast energy transfer among the particles. Here we will focus on the case $k_c\ell^2\gg k_BT$.

 In addition to the bulk properties of the model, we should stipulate the behavior at the ends of the wire. This will be done below for each particular situation.

\subsection{Definitions and statistical properties\label{defs}}
An inherent difficulty in the description of electrical variables in a circuit by means of numerical study of the particles motion is that the number of particles will be far from macroscopic, and therefore unable to reproduce the underlying assumption that current is uniform. It is therefore necessary to specify what we mean by current, voltage, and so on.
\subsubsection{Current\label{current}}
We define the instantaneous current in a resistor as its spatial average along the resistor. This gives
\begin{equation}
\tilde{I}:=(q/L)\sum v_i\;,
\label{defI}
\end{equation}
where $v_i$ is the velocity of particle $i$ and the sum is over all the particles in the resistor. The measured current $I$ is obtained by averaging $\tilde{I}$ over a lapse of time $\Delta t$. We note that (imagining the resistor as straight) the current equals the total momentum of the particles multiplied by $q/mL$.

Let us now consider a closed system of particles, as in the case that the resistor is a closed loop. Since there are no external forces, the current remains constant between collisions. Since by model assumptions $\langle v_i^2\rangle =k_BT/m$ and the velocities of the particles are uncorrelated, for $\Delta t=\tau$ we obtain $\langle I^2\rangle =Nq^2k_BT/mL^2$ and, using (\ref{Drude}), $\langle I^2\rangle =2k_BT/R_D\tau$. Finally, for measurement durations $\Delta t$ that contain several collision cycles, $\langle I^2\rangle =2k_BT/R_D\Delta t$, in agreement with (\ref{varI}).

\subsubsection{Voltage}
If the wire were perfectly homogeneous (density of particles $N/L$, independent of $x$), the electrical potential at $x=0$ relative to that at $x=L$ would be $(L/qN)\int_0^L f(x)dx$, where $f(x)dx$ is the force exerted on all the particles in a segment of length $dx$. For a limited number of particles, in order to push the current, the electric field at a position without particles is irrelevant and only the actual forces on the particles count, so the integral expression should be replaced with $(L/qN)\sum F_i$, where $F_i$ is the force exerted on particle $i$ and the sum applies to all the particles. Since the internal forces cancel out, only the external forces $F^{\rm ext}$ have to be taken into account. Finally, we note that the influence of a force on the time-averaged current is proportional to the time left until the next collision. In summary, we write
\begin{equation}
    V=(1/\Delta t)\int_0^{\Delta t} \tilde{V}(t)dt \,,
\end{equation}
with
\begin{equation}
    \tilde{V}(t):=(L/qN)(2-2t'/\tau )\sum F^{\rm ext} \,,
    \label{defV}
\end{equation}
where $t'=t-\lfloor t/\tau \rfloor \tau$ is the time elapsed since the latest collision. 

We note that the present definition of voltage deviates from the usual definition as the difference of electrical potentials. In particular, the voltage drop along two segments in series is guaranteed to equal the sum of voltage drops only in the limit that both segments have the same number of particles per length. Although the deviation from the electrical potential may appear disturbing, we should bear in mind that the quantity that voltmeters measure is not the electrical but rather the electrochemical potential, which specifies the capability to drive current.
 
\subsubsection{Units}
 Lengths, times, voltages, currents, temperatures, and $k_c$ will, respectively, be taken in units of $\ell$, $\tau$, $m\ell^2/q\tau^2$, $q/\tau$, $m\ell^2/k_B\tau^2$ and $m/\tau^2$; in cases where writing these units explicitly would be cumbersome, they will be omitted.

\subsubsection{Segments}
For any segment $PQ$, we will denote by $V_{PQ}$ the voltage at $P$ relative to $Q$, by $L_{PQ}$ the length of the segment and, in general, will use the subscript $PQ$ to indicate that a quantity refers to the segment from $P$ to $Q$. Neither the voltage nor the current can be sharply fixed in a segment, and we will study the emf $\varepsilon_{PQ} =V_{PQ}+I_{PQ}R_{PQ}$.

Since particles can enter or exit the segment, the number of particles $N_{PQ}$, and therefore the resistance $R_D$ are not fixed. When current is driven, fluctuations in the resistance are known to lead to flicker (1/f) noise \cite{Dutta,Balandin}, but in our case no current will be driven.

In the absence of a theory, we call upon a dimensional argument. From Eqs.\ (\ref{Drude}), (\ref{defI}) and (\ref{defV}), we note that
the variances of $IR_D^{1/2}$ and $VR_D^{-1}$ are independent of $N$. Motivated by this feature, we define the normalized emf
\begin{equation}
    \hat{\varepsilon}:=(1/\Delta t)\int_0^{\Delta t} [\tilde{V}(t)R^{-1}(t)R^{1/2}(\langle N\rangle)+\tilde{I}(t)R^{1/2}(t)]dt
    \label{hatemf} \,,
\end{equation}
where all the quantities refer to the same segment and $R(t)$ is obtained by setting in (\ref{Drude}) the number of particles in the segment at time $t$, whereas $R(\langle N\rangle)$ is obtained by setting the average number of particles in the segment.
If $N_{PQ}$ is constant, the Nyquist result is 
\begin{equation}
    \langle\hat{\varepsilon}_{PQ}^2\rangle\Delta t=2k_BT \,.
    \label{Nyemf}
\end{equation} In the case $\tilde{V}=0$, this result also follows from the argument provided in Section \ref{current}. However, Eq.\ (\ref{Nyemf}) is not valid in general. We will stay on the safe side by taking sufficiently small temperature and sufficiently long segments, so that the diffusion of particles into or out of the segment during the duration $\tau$ is much smaller than the number of particles that remain inside the segment.

\subsection{Results}
In our calculations, we focused on force coefficients in the range $25\le k_c\le 50$ and followed the motion of the particles between collisions with a leapfrog algorithm with time steps of size $0.01\tau$, much shorter than $\sqrt{m/k_c}$.

\subsubsection{Test of resistor behavior}
We considered a system of particles in a loop and external forces were applied to them, as would be the case if a time-dependent magnetic flux pierced the loop.

\noindent{\it Ohm's law.~}
We followed the motion of $N=20$ particles, took the values $k_c=50m/\tau^2$ and $\Delta t=10^5\tau$, and the ranges $V\le 1m\ell^2/q\tau^2$, $k_BT\le 10^3 m\ell^2/\tau^2$ and $3\ell\le L\le 40\ell$. In most cases, the same external force was applied to every particle and remained constant over time, but we also considered the case in which the external force was applied to a single particle and the case in which the force was on for a time $\tau/2$ after each collision and off during the following half of the time.

In all the cases that we examined, we obtained $V-R_DI\approx 0$, where the deviation from zero was of the order expected from Eq.\ (\ref{varV}).

\begin{figure}
\scalebox{0.5}{\includegraphics{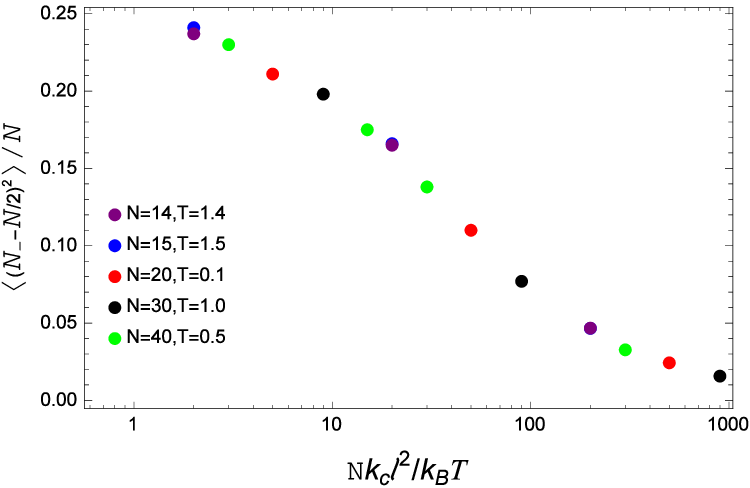}}
\caption{\label{neutral} Expected deviations from electroneutrality in a closed circuit, for several parameters. In all cases, $k_c=50m/\tau^2$, $L=10\ell$, and there was a constant emf applied by an external source, equal to $0.5m\ell^2/q\tau^2$.}
\end{figure}

\noindent{\it Electroneutrality.~}
We tested electroneutrality on a loop of length $L=10\ell$, conceptually divided into two halves. Denoting by $N_-$ the number of particles in the region $0<x\le L/2$, the deviation from electroneutrality can be characterized by the variance $\langle (N_--N/2)^2\rangle$. We evaluated this variance for several values of $N$, $T$ and $k_c$. In the limit $k_c/T\rightarrow 0$, the system becomes a gas of non-interacting particles and $\langle (N_--N/2)^2\rangle =2^{-N}N!\sum_{i=0}^N (i-N/2)^2/i!(N-i)!=N/4$. In the opposite limit, $T/k_c\rightarrow 0$, the system becomes a symmetric rigid structure (provided that $N\ell >L$), leading to $\langle (N_--N/2)^2\rangle =0$ for even $N$ and to $\langle (N_--N/2)^2\rangle =1/4$ for odd $N$. Our results are shown in Fig.\ \ref{neutral}. Our results suggest that, for all the cases considered, $\langle (N_--N/2)^2\rangle /N$ is approximately given by the same function of $Nk_c/T$. 

\subsubsection{Noise at uniform temperature}
We simulated ``measurements'' during periods of time $\Delta t=50\tau$ in the absence of applied voltage. In each measurement, the current, the voltage, and the emf were evaluated as defined in Section \ref{defs}.
To evaluate the variance, measurements were repeated $10^5$ times. Before the measurements, the system typically evolved for $2\times 10^5\tau$ to ensure stabilization, and idle times of $10\tau$ were left between consecutive measurements to avoid correlation. For the parameters we used, the influence of the factor $2-2t'/\tau$ in definition (\ref{defV}) and that of $R^{-1}(t)R^{1/2}(\langle N\rangle)$ rather than just $R^{-1/2}(t)$ in definition (\ref{hatemf}) was marginal.

\begin{figure}
\scalebox{0.5}{\includegraphics{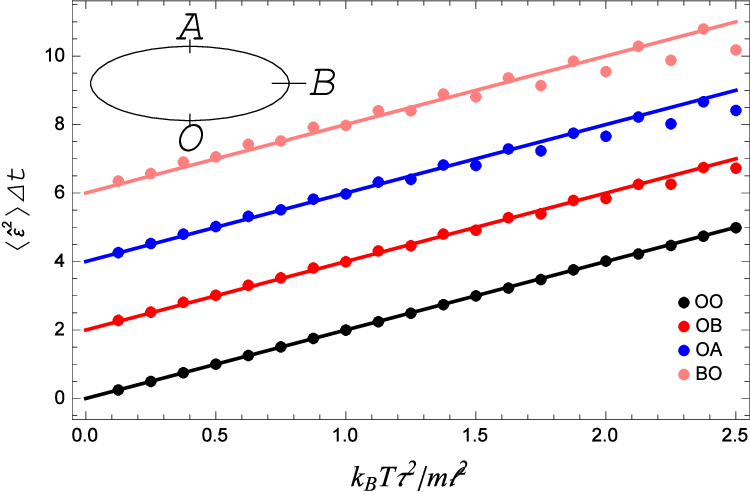}}
\caption{\label{varT}Variances of the normalized efm [as defined in Eq.\ (\ref{hatemf})] in a closed circuit, as functions of the temperature. As shown in the inset, $OO$ ($OB$, $OA$, $BO$) has length $L$ ($3L/4$, $L/2$, $L/4$), where $L=40\ell$. The points for which the temperature is an odd (even) multiple of 0.125 were obtained setting $k_c=50m/\tau^2$ ($k_c=25m/\tau^2$); for other parameters, see text. For visibility, the results for different segments have been shifted upwards in steps of 2 units. The straight lines show the values predicted by Eq.\ (\ref{varep}). Since all the points in the circuit are equivalent, $\langle\hat{\varepsilon}\rangle =0$ for every segment. }
\end{figure}

\noindent{\it Closed circuit.~} We studied a loop with $N=50$ particles and length $L=40\ell$ for the cases $k_c=25$ and $k_c=50$, and evaluated the variance of the normalized emf in the entire loop, and also in the segments described by the inset in Fig.\ \ref{varT}. Figure \ref{varT} shows these variances as functions of the temperature.

We see that for $k_BT\lesssim 1.5m\ell^2/\tau^2$ our results are in good agreement with Eq.\ (\ref{varep}), but for the smaller value of $k_c$ and $k_BT\agt 1.5m\ell^2/\tau^2$ there is a systematic deviation, that increases as the length of the segment decreases. We attribute the deviation from Eq.\ (\ref{varep}) for small values of $Nk_c/T$ to the deviation from electroneutrality, that is a built-in assumption in the Nyquist scenario.

\begin{figure}
\scalebox{0.5}{\includegraphics{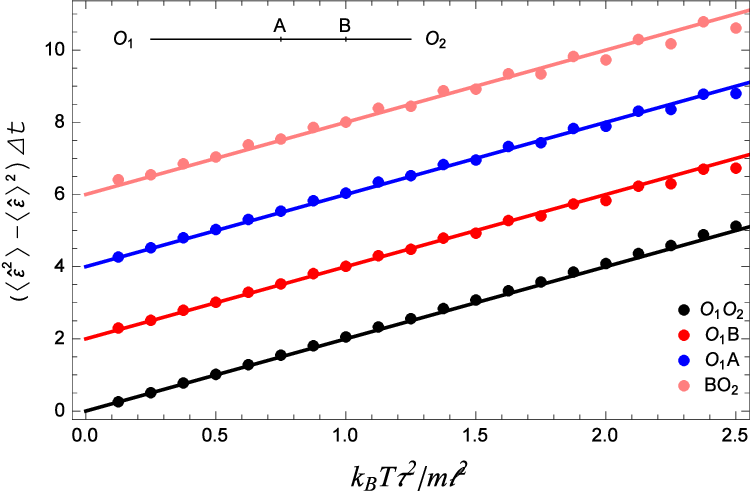}}
\caption{\label{unifopen}Variances of the normalized efm [as defined in Eq.\ (\ref{hatemf})] in an open circuit, as functions of the temperature. As shown in the inset, $O_1O_2$ ($O_1B$, $O_1A$, $BO_2$) has length $L$ ($3L/4$, $L/2$, $L/4$), where $L=40\ell$. All the parameters are the same as in Fig.\ \ref{varT}. For visibility, the results for different segments have been shifted upwards in steps of 2 units. The straight lines show the values predicted by Eq.\ (\ref{varep}). }
\end{figure}

\noindent{\it Open circuit.~} We open the circuit considered above by cutting the wire at the point $O$ and creating two ends, $O_1$ and $O_2$, as shown in the inset of Fig.\ \ref{unifopen}.

Some force is required at the ends to counter the repulsion from the internal particles and thus prevent the escape of particles from the wire. With the purpose of having similar environments for those particles close to an end and those in the bulk, we impose that for every particle at position $x$ there are mirror charges at $-x$ and $2L-x$, and these charges exert the same repulsion force as a real particle. If, despite this force, a particle reaches an end, we assume that it bounces elastically.

Our results are shown in Fig.\ \ref{unifopen}. Again, there is good agreement with Eq.\ (\ref{varep}) for $k_BT<1.5m\ell^2/\tau^2$. 

\subsubsection{Circuits with resistors at different temperatures\label{T1neqT2}}
In the case of an open circuit with a limited number of particles and $T=0$, the particles assume equilibrium positions and form a lattice. For $0<k_BT\ll k_c\ell^2$, there is still a lattice and the particles vibrate about their equilibrium positions, so that space is not homogeneous. As a consequence, the results are expected to depend on the particular places where the temperature changes and the particular locations of the ends of any considered segment. 
Inhomogeneity of space may become negligible for sufficiently long segments; this question will be postponed for a future study, and here we focus on closed circuits. 

\begin{figure}
\scalebox{0.5}{\includegraphics{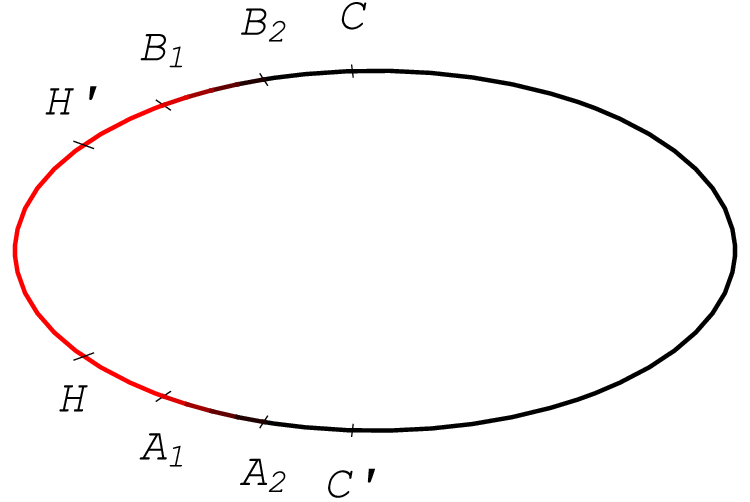}}
\caption{\label{CH} Closed circuit with nonuniform temperature. Segment $A_1B_1$ ($B_2A_2$) has temperature $T_1$ ($T_2$); in $A_2A_1$ and $B_1B_2$ the temperature changes gradually.}
\end{figure}

The circuit considered is sketched in Fig.\ \ref{CH}. Region 1, from $A_1$ to $B_1$, is held at temperature $T_1$, and region 2,  from $B_2$ to $A_2$, is held at temperature $T_2$. Along segments $A_2A_1$ and $B_1B_2$ the temperature changes gradually; for numerical convenience, we take $T^{1/2}$ as a linear function of position. Segment $HH'$ ($CC'$) has temperature $T_1$ ($T_2$).

Our results are reported in Table \ref{tab:T1025}. Unless stated otherwise, we took $T_1=1$, $T_2=0.25$, $\Delta t=50$, $N=50$, $L_{A_2A_1}=L_{B_1B_2}=2$, $L_{A_1H}=L_{H'B_1}=L_{B_2C}=L_{C'A_2}=1$, $L_{HH'}=10$, and $L_{CC'}=22$, so that the sum of the segment lengths is $L=40$. We emphasize that segments $HH'$ and $CC'$ have uniform temperatures. 
By symmetry, $\langle\varepsilon_{HH'}\rangle =\langle\varepsilon_{CC'}\rangle =\langle V_{HH'}\rangle =\langle V_{CC'}\rangle =0$ and $\langle\varepsilon_{C'H}^2\rangle -\langle\varepsilon_{C'H}\rangle^2=\langle\varepsilon_{H'C}^2\rangle -\langle\varepsilon_{H'C}\rangle^2$.

\noindent{\it Nyquist scenario} (NS).~ The variances of $\varepsilon_{HH'}$ and $\varepsilon_{CC'}$ were evaluated using Eq.\ (\ref{varep}), and those of $\varepsilon_{H'C}$ and $\varepsilon_{C'H}$, using Eq.\ (\ref{steady1}). The variance of $V_{PQ}$ is obtained noting that $V_{PQ}=\varepsilon_{PQ}-IR_{PQ}=(\varepsilon_{PQ}\sum 'R_{P'Q'}-R_{PQ}\sum '\varepsilon_{P'Q'})/R$, where $\sum '$ denotes sum over all the segments except $PQ$ and $R$ is the resistance of the entire circuit. From here, Eq.\ (\ref{varep}), and the assumption that the emf in different segments are uncorrelated, 
\begin{equation}
\frac{S( V_{PQ})}{R_{PQ}}=\frac{S(\varepsilon_{PQ})(\sum 'R_{P'Q'})^2/R_{PQ}+R_{PQ}\sum 'S(\varepsilon_{P'Q'})}{R^2} \,
\label{eptoV}
\end{equation}
where $S(X)$ is the variance of $X$. For our evaluations in the NS we have assumed that the resistivity is uniform.

We note that Eq.\ (\ref{eptoV}) applies to a closed circuit; in an open circuit, $I\approx 0$ and $V_{PQ}\approx\varepsilon_{PQ}$. We also note that for a uniform wire in the limit $R_{C'H,H'C}\ll R_{HH',CC'}$, $S(V_{HH'})R_{CC'}/S(V_{CC'})R_{HH'}=T_1/T_2$, so that in this limit, Eq.\ (\ref{eptoV}) coincides with Eq.\ (\ref{varratio}).

\noindent{\it Limit} $k_c\rightarrow\infty$.~ This is the case of a rigid structure analyzed in Appendix \ref{strong}. In order to have values comparable with those of our simulations, while complying with the assumptions of Appendix \ref{strong}, we took $L_{C'H}=L_{H'C}=0$ and $N_1/L_{HH'}=N_2/L_{CC'}=\mathrm{integer}$ (as required by the assumptions in the Appendix), and $L_{CC'}/L_{HH'}=2.2$ (as the ratio in the most similar cases in the simulation). The variances of $\varepsilon_{PQ}$ follow from Eq.\ (\ref{rigidepsq}), and those of $V_{PQ}$, from Eq.\ (\ref{rigidVsq}). Note that, counterintuitively, in this extreme case the variance of the emf in the colder segment can be larger than that in the hot segment.

\noindent{\it Large} $k_c$.~ This and the following rows were obtained from our simulations. This case is intended to imitate the situation studied in Appendix \ref{strong}. We therefore require $L_{H'C}=L_{C'H}=0$ (which implies a discontinuous passage between $T_1$ and $T_2$) and integer value of $N_1/L_{HH'}=N_2/L_{CC'}$. We took $N=N_1+N_2=48$, $L_{HH'}=7.5$ and $L_{CC'}=16.5$, so that $N_1/L_{HH'}=N_2/L_{CC'}=2$. As a large value of $k_c$ we took $k_c=100$ and, accordingly, small evolution steps every $\tau /150$. 

\noindent{\it Cases} $k_c=50$ {\it and} $k_c=25$.~ We see that the variances $S(\varepsilon_{HH'})$ and $S(\varepsilon_{CC'})$ strongly differ from the NS, and assume intermediate values between the NS and the rigid structure case. Remarkably, the normalized variance is larger in the cold region than in the hot region, as in the stromg-interaction limit.

\begin{table*}
\caption{\label{tab:T1025} Variances of the emf in some of the segments shown in Fig.\ \ref{CH}. The first two rows are the predictions of two competing models and the following rows are simulation results. 
NS stands for the Nyquist scenario and the case $k_c\rightarrow\infty$ is studied in Appendix \ref{strong}. Since the number of particles in each segment is not fixed, the variances were evaluated using (\ref{hatemf}). For additional parameters and details, see Sec.\ \ref{T1neqT2}. The horizontal lines are guide for the eye.}
\begin{ruledtabular}
\begin{tabular}{lccccc}
case & $\frac{\langle\varepsilon_{HH'}^2\rangle\Delta t }{R_{HH'}}$&
$\frac{\langle\varepsilon_{CC'}^2\rangle\Delta t }{R_{CC'}}$& $\frac{(\langle\varepsilon_{C'H}^2\rangle -\langle\varepsilon_{C'H}\rangle^2)\Delta t}{R_{C'H}}$
&$\frac{\langle V_{HH'}^2\rangle R_{CC'}}{\langle V_{CC'}^2\rangle R_{HH'}}$ &
$\frac{N_{HH'}L_{CC'}}{N_{CC'}L_{HH'}}$\\
\hline
NS & 2.00 & 0.50 &1.21
&2.46&\\
$k_c\rightarrow\infty$, $L_2/L_1=2.2$ & 0.71 & 1.79 &
&$2.20$&1~~~\\
large $k_c$, $L_2/L_1=2.2$ & 1.00 & 1.82 &
&2.23&0.99\\
$k_c=50$, $N=50$ & 1.25 & 1.52 &1.32
&2.66&0.99\\
\hline
$k_c=25$, $N=50$ & 1.12 & 1.46 &1.24
&2.50&0.96\\
$k_c=25$, $N=100$ &1.13  & 1.22 &1.29 
&2.40 &0.96 \\
$k_c=25$, $N=200$ & 1.42 & 0.87 &1.30
&2.39&0.95\\
$k_c=25$, $N=400$ & 1.70 & 0.67 &1.24
&2.35&0.95\\
\hline
$k_c=10$, $N=200$  & 1.64 & 0.66 &1.00
&2.19&0.84\\
$\rho_1\neq\rho_2$, $k_c=25$ & 1.11 & 1.53 & 
&2.47&0.96\\
perfect, $k_c=25$ & 0.98 & 1.55 & 
&2.38&0.96\\
perfect, $k_c=10$ & 1.64 & 0.66 & 
&2.02 &0.85\
\end{tabular}
\end{ruledtabular}
\end{table*}

\noindent{\it Cases} $N=100,\,200,\,400$.~ As seen above, the value $k_c=25$ leads to a very rigid structure. We would like to study softer structures, but taking $k_c$ below 25 would confine the validity of our simulations to lower temperatures. Instead, we keep $k_c=25$ and take longer circuits. $L_{H'C}$ and $L_{C'H}$ remain unchanged, but for $N=100$ (200, 400) $L_{HH'}=23$ (49, 101), $L_{CC'}=49$ (103, 211), so that $L_{CC'}/L_{HH'}\approx 2.1$ and $L=0.8N$. Indeed, as the length of the circuit increases, $S(\varepsilon_{HH'})$ and $S(\varepsilon_{CC'})$ become closer to the NS values. We consistently find that $S(\varepsilon_{HH'})$ [$S(\varepsilon_{CC'})$] is smaller [larger] than predicted by the NS, and, for $N\agt 100$, the deviations from the NS values are roughly inversely proportional to $N$. Our results suggest that the presence of a resistor at a different temperature may be felt at distances (from center to center) that exceed the screening length by three orders of magnitude.

For comparison, we sometimes repeated our simulations setting $T_2=T_1=1$. For example, in the case $N=200$ we obtained $S(\hat\varepsilon_{HH'})\Delta t=1.91$ and $S(\hat\varepsilon_{CC'})\Delta t=1.95$; both results differ from the NS prediction by less than $5\%$. We also evaluated the variance of $V_{\rm total}=V_{HH'}+V_{H'C}+V_{CC'}+V_{C'H}$, which ideally should vanish; $S(V_{\rm total})\Delta t$ came out to be smaller than $Rk_BT_{1,2}$ by two orders of magnitude.

\noindent{\it Case} $k_c=10$.~ For longer circuits, and accordingly longer $L_{HH'}$ and $L_{CC'}$, we can take smaller values of $k_c$. The lengths of the segments in this case were the same as in the case ``$k_c=25$, $N=200$". Comparison of the rows ``$k_c=25$, $N=200$,'' ``$k_c=25$, $N=400$,'' and ``$k_c=10$, $N=200$'' indicates that increasing the length of the circuit has an effect similar to that of decreasing the value of $k_c$.

\noindent{\it Resistors made of different materials} ($\rho_1\neq\rho_2$).~ At uniform temperature, the variances of the emf depend on the resistances but do not depend separately on the resistivity and geometry of each resistor. As a test for the extension of this property to nonuniform temperature, we considered a circuit with two parts. Part $i$ had resistance $\rho_i$ per unit length and the central segment in it had temperature $T_i$. We took $\rho_2=2\rho_1$, $k_c=25$, and the number of particles and lengths were chosen to produce resistances similar to the case ``$k_c=25$, $N=50$''. The resistivity $2\rho_1$ was achieved by means of collisions every lapse of time $\tau /2$ rather than $\tau$. The gradient of $T^{1/2}$ was taken proportional to the resistivity and the temperature changes between $T_1$ and $T_2$ took place over the distance $1.5\ell$. To keep the same average density as in the case ``$k_c=25$, $N=50$'', we took $L_{A_1H}=L_{H'B_1}=1$, $L_{B_2C}=L_{C'A_2}=0.5$, $L_{HH'}=10$, $L_{CC'}=11.2$ and $N=1.25L=34$. The results we obtained, as shown in Table \ref{tab:T1025}, are similar to those of case ``$k_c=25$, $N=50$''. The voltages and emf in segments $H'C$ and $C'H$, where the wire is not uniform, are not evaluated.

\noindent{\it Resistors connected by perfect conductors} (``perfect'').~ This situation is simulated by replacing the regions with temperature gradient considered above with regions where there are no collisions (and therefore the lattice temperature in these regions is irrelevant). We considered two cases: one almost identical to the case ``$k_c=25$, $N=50$'' and the other almost identical to the case `$k_c=10$, $N=200$''; the only difference from the above cases is that segments $A_2A_1$ and $B_1B_2$ became collisionless. We note that in the present cases there is a slight tendency (only for the shorter circuit) for a larger departure from the NS values than in the corresponding former cases, suggesting that the effective interaction between the hot and the cold regions is transmitted more freely along a perfect conductor than along a resistive wire.

\subsubsection{Influence of the temperature gradient}
\noindent{\it Heat flow}.~ A temperature gradient causes heat flow, and Sukhorukov and Loss \cite{Loss} found that heat flow can be a source of electric noise. Their result relates the noise at a given terminal to the thermal currents through all the terminals, but we may expect that the larger the local thermal current, the stronger its influence in its vicinity. As a check for this prediction, we evaluated the variances of $\varepsilon_{C'H}$ and $\varepsilon_{H'C}$, but the short length over which the temperature changes makes it hard to draw any conclusion. 

\begin{figure}
\scalebox{0.5}{\includegraphics{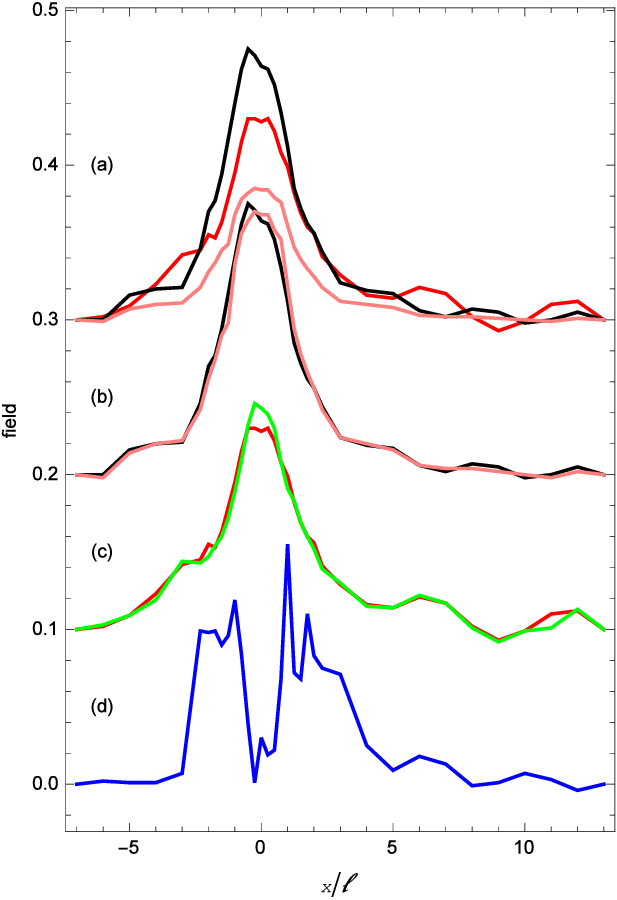}}
\caption{\label{Sfield} Analogue of the electric field induced by the nonuniform temperature, in a half of a symmetric closed circuit. The origin $x=0$ was taken at the middle of $A_2A_1$, so that the temperature is uniform for $|x|>\ell$. For visibility, panel (a) [(b), (c)] was raised by 0.3 [0.2, 0.1] units. Red: $k_c=50$, $T_2=0.25$; black: $k_c=25$, $T_2=0.25$; pink: $k_c=25$, $T_2=0.625$; green: $k_c=25$, $T_2=0.25$, smooth temperature; blue: $k_c=25$, $T_2=0.25$, resistors joined by perfect conductors. See text for additional parameters.}
\end{figure}

\noindent{\it Seebeck effect} \cite{Lee}.~ Since diffusivity increases with temperature, a force from lower to higher temperature is required to maintain a steady state. We define ``field" at position $x$ (in analogy to electric field) as the time-average of the force applied on the particles in the vicinity of $x$, divided by $q$. This field was evaluated with our simulations by letting them run during $1.4\times10^6\tau$ for stabilization, and then taking the average during $5.6\times10^6\tau$. We studied the circuits with $N=50$ described in Sec.\ \ref{T1neqT2} and took $T_1=1$. In Fig.\ \ref{Sfield} we present the field as a function of $x$ in half of the circuit, between the middles of $HH'$ and $CC'$. The origin $x=0$ was taken at the middle of $A_2A_1$, leading to the range $-7\le x\le 13$. By symmetry of the circuit, the field at $26-x$ equals minus the field at $x$.

The case $k_c=25$ ($k_c=50$) and $T_2=0.25$ is depicted by the black (red) line in panel (a) of Fig.\ \ref{Sfield}. The pink line corresponds to $k_c=25$ and $T_c=5/8$, i.e., the value of $T_1-T_2$ is half that of the other lines. We note that the Seebeck field is present not only in the region $|x|<\ell$ in which the temperature is not uniform, but extends several screening lengths beyond $A_2A_1$. We also note that despite the larger interparticle force in case $k_c=50$, it results in fields smaller than in case $k_c=25$; the reason is that $k_c$ imposes a more uniform density of particles.

From the Seebeck effect we expect that the integral of the field be proportional to $T_1-T_2$. Panel (b) of Fig.\ \ref{Sfield} shows the line for $k_c=25$ and $T_2=2/8$ (black) together with the line for $k_c=25$ and $T_2=5/8$ multiplied by a factor 2 (pink). These lines almost coalesce, indicating that the proportionality holds not only for the integrals, but locally for the fields. 

The jaggedness of the lines is due to the discontinuity of the temperature gradient at $x=\pm\ell$. In panel (c) of Fig.\ \ref{Sfield} the red line stands for our usual case $k_c=50, T_2=0.25$, whereas the green line corresponds to the same values, but in the regions $A_2A_1$ and $B_1B_2$ $T^{1/2}$ does not depend linearly on position, but rather as a cubic polynomial such that $dT/dx$ vanishes at $x=\pm\ell$. The remaining jaggedness of the green line is mainly due to the relatively large distance between consecutive evaluation points. 

A smoother function $T(x)$ leads to a smoother dependence of the field on position, but does not suppress the oscillations that are conspicuous near the middle of the cold segment. We conjecture that these oscillations are due to a standing acoustic wave, that is excited by the hot region. Support for this conjecture is provided by the fact that the relative size of these oscillations decreases when $T_1-T_2$ becomes smaller (more precisely, the size of the oscillations decreases by a larger factor than the spatial average of the field). An example of this behavior can be observed in panel (b) of Fig.\ \ref{Sfield}.

Panel (d) of Fig.\ \ref{Sfield} takes again $k_c=25$ and $T_2=0.25$, but this time $A_2A_1$ and $B_1B_2$ are perfect conductors. As might naively be expected, the field within a perfect conductor is small, and peaks at $|x|\ge\ell$. 

\subsubsection{Correlations\label{seccorr}}
We denote the correlation between two random variables $X_{1,2}$ with zero average by $r(X_1,X_2):=\langle X_1 X_2\rangle /\sqrt{S(X_1)S(X_2)}$; for consistency, $X_1$ and $X_2$ are understood as time averages of the corresponding physical quantities during common lapses in time of size $\Delta t$.
We consider a closed circuit as in Fig.\ \ref{CH}. In the following of this Section, quantities that refer to segment $HH'$ ($CC'$), which are kept at temperature $T_1$ ($T_2$), will be marked with the index 1 (2). 

The usual circuit theory predicts several relations. Electroneutrality implies $r(I_1,I_2)=1$. The NS predicts $r(\varepsilon_1,\varepsilon_2)=0$ and, for $T_1=T_2$, thermodynamics leads to $r(V_1,I_1)=r(V_2,I_2)=0$. If segments $H'C$ and $C'H$ are perfect conductors or have vanishing length, then $r(V_1,V_2)=-1$.

We have investigated to what extent our simulations comply with the above predictions. Taking $N=50$, $L_1=10$, $L_2=22$, $L_{C'H}=L_{H'C}=4$, $L_{A_2A_1}=L_{B_1B_2}=2$ with $A_2A_1$ and $B_1B_2$ perfect conductors, fixing the average temperature  $(T_1+T_2)/2=0.6$, and varying the temperature difference $T_1-T_2$, we obtained the results as follows. $r(I_1,I_2)$ varies between 0.93 at $T_1-T_2=-1.2$ and 0.84 at $T_1-T_2=1.2$; the deviation from 1 is expected because electroneutrality is not perfect. $r(V_1,V_2)$ varies between -0.64 at $T_1-T_2=-1.2$ and -0.78 at $T_1-T_2=1.2$; the deviation from -1 is expected because segments $H'C$ and $C'H$ have resistance. The differences between the cases $T_1-T_2=-1.2$ and $T_1-T_2=1.2$ are expected because $HH'$
and $CC'$ have different lengths.

\begin{figure}
\scalebox{0.5}{\includegraphics{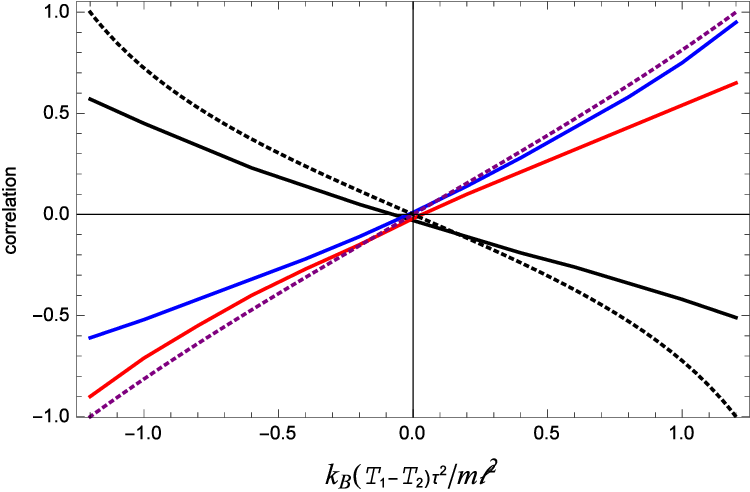}}
\caption{\label{correlations} Correlations between the emf's in two detached segments and between the voltage and the current in the same segment, as functions of the temperature difference. The dotted line corresponds to the strong interaction limit and the solid line to $k_c=50$. Black: $r(\varepsilon_1,\varepsilon_2)$; red: $-r(V_1,I_1)$; blue: $r(V_2,I_2)$; purple: Eq.\ (\ref{corrVI}). The average temperature is $(T_1+T_2)/2=0.6$. For additional parameters see Sec.\ \ref{seccorr}. We attribute the deviations of $r(V_1,I_1)$ and $r(V_2,I_2)$ from 0 for $T_1=T_2$ to inaccuracy of our algorithm.} 
\end{figure}

$r(\varepsilon_1,\varepsilon_2)$ and $r(V_i,I_i)$ are shown in Fig.\ \ref{correlations}. The solid lines correspond to the present case and the dotted lines to the strong interaction limit (Appendix \ref{strong}). $r(\varepsilon_1,\varepsilon_2)$ is a nearly odd function of the temperature difference and behaves qualitatively as in the strong interaction limit. The fact that $r(\varepsilon_1,\varepsilon_2)\neq 0$, even for detached segments, invalidates a central assumption on which Eq.\ \ref{steady1} is based. Denoting by $j$ the segment that is not $i$, $r(V_i,I_i)$ is positive (negative) if $T_i<T_j$  ($T_i>T_j$). This is expected because $\langle V_iI_i\rangle$ is the rate of work performed by segment $j$ on the center of mass \cite{pseudo} of the particles in segment $i$, and the hotter segment has to deliver energy to the colder segment. $|r(V_i,I_i)|<|r(V_j,I_j)|$ for $T_i>T_j$ because for lower temperature the particles behave as a more rigid structure and a higher fraction of their kinetic energy can be ascribed to their center of mass; according to this argument, Eq.\ (\ref{corrVI}) sets an upper bound on $|r(V_i,I_i)|$.

\subsection{Nonstationary situations}
Our model is insufficient for the study of situations in which temperatures vary in time, because not only the charges, but also the medium in which they are located, contribute to the internal energy and to heat transport. 

In order to have a meaningful concept of temperature (even local temperature), the typical times for temperature changes have to be much longer than $\tau$. 

Since the time required by a circuit to approach thermal equilibrium does depend on the calorimetric properties of the medium, whereas the time required for propagation of an electric signal along the wire does not depend on these properties, a circuit can be engineered to allow the measurement proposed in Fig.\ \ref{verify}. If the initial temperatures of the resistors in Fig.\ 2 are well defined and if the time required for charge equilibration is much shorter than the time required for heat equilibration, then, immediately after charge equilibration, we should recover the results obtained for the stationary situation.

\section{Discussion and perspective}

Since Johnson's measurement of thermal noise in a resistor and the immediate thermodynamic explanation by Nyquist, this subject has been regarded as closed. Even the case of nonuniform temperature, for which the thermodynamic argument does not apply, is generally taken into account assuming that the medium can be divided into pieces, each with a definite local temperature, thus extending the Nyquist formula to Eq.\ (\ref{steady1}).


Although the macroscopic behavior of the Johnson noise is well understood, we suggest that the microscopic mechanism for its emergence still requires investigation. 
Different microscopic mechanisms could be distinguished in circuits at nonuniform temperatures, in which the interparticle interaction could lead to spatial correlation and non-local temperature dependence.

The available experimental results hint at a nonlocal interplay between resistors in contact, which leads to deviations from Eq.\ (\ref{steady1}), but these deviations are not substantial and do not manifest themselves unless the experimenter looks for them. We have suggested experiments in which these deviations would be enhanced.

Our simulations are intended to provide an additional source of motivation to look for the effects that we suggest. The main limitations of our simulations are that (1) they are strictly one-dimensional, probably involving correlation properties that are not representative of real thin wires; as any simulation, they are based on a limited number of particles, and therefore (2) the definitions of current and voltage, which are appropriate for a continuum of charge, require extension to the discrete case, and (3) charge variations in a segment, due to thermal inhomogeneity or fluctuations, can be a significant fraction of the charge in the segment. 
The particles in our model are classical objects, but due to the very wide scope of Nyquist's analysis, this is not expected to be a limitation for $\Delta t>\hbar /k_BT$.
Future projects could use the model introduced here to deal with systems such as (1) open circuits, in which the emf is more closely related to the voltage, which is directly measurable, (2) nets of wires and junctions, that may relax features that arise from one-dimensionality, (3) two-dimensional wires, in which the lateral walls induce image charges and reflections, as we assumed at the ends of an open circuit, (4) system with weak interparticle interaction, which could hopefully be treated analytically and provide a complementary limit to that obtained in Appendix \ref{strong}, or (5) wires in which the lattice has time dependent temperature as a result of heat diffusion. 

The results of Sec.\ \ref{T1neqT2} suggest that the presence of a resistor with different temperature could be felt at a distance along the circuit of the order of $10^3$ times the screening length. For instance, if this result is applicable to strontium titanate \cite{AR,PNAS} with dopant concentration of $10^{17}$cm$^{-3}$, at temperatures in the range $25-100\,$K, this distance would be of the order of a tenth of a millimeter.

In summary, the available experimental and theoretical results that we have reviewed, as well as our simulation results, hint at a non-local interplay between resistors in electrical contact, altering the predictions of the Nyquist scenario. The conceptual significance of this result is that, unless measuring circuits offset non-local influence, \cite{Brian} corrections may be required in Johnson noise thermometry.
The technological significance of the scenario conjectured here is the possibility of reducing thermal noise in a part of a conductor by lowering the temperature in another part of it.

\section*{Acknowledgments}
We thank Antoine B\'{e}rut, Frederick Green, Mordehai Heiblum, Yariv Kafri, Igor Khovanov, Michael Reznikov, and Andrey Varlamov for useful advice. This study was partially supported by the Israel Science Foundation (ISF).  

\appendix
\section{Johnson noise in the time domain\label{Brown}}
For simplicity, let us consider a uniform resistor in the shape of a wire of length $L$ and cross section $A$, shorted to enforce $V=0$. Let the flowing charges be a gas of particles with charge $q$ and volume density $n$, and let $\Delta t$ be sufficiently larger than the average time between collisions, so that Brownian motion of the flowing charges can be described by the diffusion equation.

If a particle moves a short distance $x$ along the wire, it will result on the average in the passage of a charge $\Delta Q_1=qx/L$ across an arbitrary surface that cuts the wire. From the diffusion equation we know that $\langle x^2\rangle =2D\Delta t$, where $D$ is the diffusion coefficient, and therefore the variance of the charge passed by one particle is
\begin{equation}
    \langle (\Delta Q_1)^2\rangle =2q^2 D\Delta t/L^2 \,.
    \label{DQ1}
\end{equation}
There are $nAL$ charged particles in the gas. Assuming that their motions are uncorrelated, the variance $\langle (\Delta Q)^2\rangle$ of the total charge that passes the surface will be $nAL\langle (\Delta Q_1)^2\rangle$, i.e.
\begin{equation}
    \langle (\Delta Q)^2\rangle =2nAq^2 D\Delta t/L \,.
    \label{DQ}
\end{equation}
Invoking the Einstein relation $D=k_BT\mu /q$, \cite{ein,Reif} where $\mu$ is the mobility, noting that $R=L/nq\mu A$, and dividing (\ref{DQ}) by $(\Delta t)^2$, we recover Eq.\ (\ref{varI}).

Equation (\ref{varV}) follows from (\ref{varI}) by noting that, if the current in the absence of voltage is $\eta_I\sqrt{2k_BT/R\Delta t}$, then the voltage required to cancel this current is $-\eta_I\sqrt{2k_BTR/\Delta t}$. More generally, the emf required to induce a current $I$ is $RI-\eta_I\sqrt{2k_BTR/\Delta t}$, leading to Eq.\ (\ref{varep}).

On passing from (\ref{DQ1}) to (\ref{DQ}) we assumed that the motions of the charged particles are uncorrelated. This condition may not be fulfilled, because charges cannot accumulate. Nevertheless, the validity of (\ref{DQ}) may be more general than the assumed requirement. Let us consider the opposite situation, in which the charged particles move in a stiff formation. This situation is actually equivalent to that of a single particle, in which both the charge and the mass are larger by a factor $nAL$ than in the case of the gas. Since $D$ is inversely proportional to the mass of the particle, (\ref{DQ}) remains unchanged.

For $\Delta t\gg\hbar /k_BT$, Eq.\ (\ref{varV}) follows also from the Nyquist expression in the frequency domain.
In the notation of Ref.\ \cite{Reif}, the spectral density of the fluctuating voltage is $J_+(\omega )=2k_BTR/\pi$ for any angular frequency $\omega\ll k_BT/\hbar$. Therefore, the autocorrelation $\langle V(t)V(t+s)\rangle$ is
\begin{equation}
 K(s)=(2k_BTR/\pi)\int_0^\Omega \cos\omega s\,d\omega=(2k_BTR/\pi s)\sin\Omega s\;,
 \label{Ks}
\end{equation}
where $\Omega$ is a cutoff.

The average voltage measured during a period of time $\Delta t$ is 
$(1/\Delta t)\int_{-\Delta t/2}^{\Delta t/2}V(t)dt=(1/\Delta t)\int_{-\Delta t/2-t}^{\Delta t/2-t}\,V(t+s)ds$, 
so the ensemble average of its square is

\begin{align} 
\langle V^2\rangle &=\frac{1}{(\Delta t)^2}\left\langle \int_{-\Delta t/2}^{\Delta t/2}V(t)dt\int_{-\Delta t/2-t}^{\Delta t/2-t}V(t+s)ds \right\rangle\\ &=\frac{1}{(\Delta t)^2}\int_{-\Delta t/2}^{\Delta t/2}dt\int_{-\Delta t/2-t}^{\Delta t/2-t}ds\,K(s)\,.
\end{align}

Using (\ref{Ks}), evaluating the integrals, and taking the limit $\Omega\rightarrow\infty$, we recover Eq.\ (1).

\section{Strong interaction limit\label{strong}}
We study a loop of length $L=L_1+L_2$, where the region with length $L_i$ is kept at temperature $T_i$ (with discontinuous temperature rather than regions of gradual variation).
We assume that the particles are kept at fixed interparticle distances $L/N$ by rigid rods, and assume further that $N_i=NL_i/L$ are integer. This feature enables us to know exactly the number of particles in region $i$, namely, $N_i$. Every lapse of time $\tau$, after absorbing the momentum from the particles, the lattice gives an impulse $\Sigma_i$ to the particles in region $i$, with variance
\begin{equation}
   \langle\Sigma_i^2\rangle =N_imk_BT_i\,. 
   \label{varSig}
\end{equation}
The velocity acquired by the structure is $(1/Nm)(\Sigma_1+\Sigma_2)$ and the current (the same in both regions) is $I=(q/Lm)(\Sigma_1+\Sigma_2)$.

The momentum acquired by the particles in region 1 is $(N_1/N)(\Sigma_1+\Sigma_2)$. $\Sigma_1$ was provided by the lattice in the region, and the remaining $(N_1/N)\Sigma_2-(N_2/N)\Sigma_1$ is due to external forces. Using (\ref{defV}), integrating over the lapse of time $\tau$, noting that the force is applied at the beginning of this lapse of time and that the density of particles is uniform, the average voltage in region 1 is $V_1=(2L/qN^2\tau)(N_1\Sigma_2-N_2\Sigma_1)$. We note that $V_2=-V_1$ follows from Newton's third law.

With the expressions above for the current and the voltage, and with the resistance $R_1=2mL^2N_1/q^2N^2\tau$ obtained from (\ref{Drude}), we can evaluate the average emf $\varepsilon_1=V_1+R_1I$ in region 1 during the lapse of time $\tau$:
\begin{equation}
    \varepsilon_1=(2L/qN^2\tau)[(N_1-N_2)\Sigma_1+2N_1\Sigma_2] \,.
    \label{rigidep}
\end{equation}

Finally, using (\ref{varSig}), and for $\Delta t/\tau$ collision cycles, we obtain the variances
\begin{equation}
    \langle V_1^2\rangle =(2R_1N_2k_B/N^2\Delta t)(N_1T_2+N_2T_1)\,,
    \label{rigidVsq}
\end{equation}

\begin{equation}
    \langle\varepsilon_1^2\rangle =(2R_1k_B/N^2\Delta t)[(N_1-N_2)^2T_1+4N_1N_2T_2]\,,
    \label{rigidepsq} 
\end{equation}
and the analogous relations hold for region 2. For $T_1=T_2$, Eq.\ (\ref{rigidVsq}) leads to (\ref{etaij}). 
A remarkable consequence of Eq.\ (\ref{rigidVsq}) is that the ratio
\begin{equation}
   \langle V_1^2\rangle R_2/\langle V_2^2\rangle R_1=N_2/N_1
   \label{varratio}
\end{equation}
is independent of the temperatures.
The most intriguing case in Eq.\ (\ref{rigidepsq}) is that of $N_1=N_2$, in which the Johnson noise does not depend on the local temperature, but rather on that of the neighboring region.

Noting that the correlation $r(\varepsilon_1,\varepsilon_2)=\langle\varepsilon_1\varepsilon_2\rangle /\sqrt{S(\varepsilon_1)S(\varepsilon_2)}$ does not depend on $\Delta t$, from (\ref{rigidep}) and (\ref{varSig}) we also obtain
\begin{equation}
    r(\varepsilon_1,\varepsilon_2)=\frac{\beta (T_1-T_2)}{\sqrt{(\beta^2 T_1+T_2)(T_1+\beta^2 T_2)}}\,,
\end{equation}
with $\beta =(N_1-N_2)/2\sqrt{N_1N_2}$. This correlation may be either positive or negative, depending on whether the hotter segment is the longer or the shorter segment. For $T_1=T_2$, we recover the NS.

Similarly, from the expressions for $I$ and $V_1$ we obtain
\begin{equation}
    r(V_1,I)=\frac{T_2-T_1}{\sqrt{(N_1 T_1+N_2T_2)(N_2 T_1+N_1T_2)/N_1N_2}}\,.
    \label{corrVI}
\end{equation}
We note that $V_1 I$ is the power passed to region 1 by the forces that act from region 2; in the present limit, the heat transferred by the passage of particles vanishes. Note also that $r(\varepsilon_1,\varepsilon_2)$ and $r(V_1,I)$ are antisymmetric under the exchange of $T_1$ and $T_2$, and their absolute values reach 1 when one of the temperatures vanishes.


\begin{thebibliography}{9}
\bibitem{shot} Shot noise was required for this purpose.
\end{thebibliography}
\end{document}